\documentclass[twocolumn,showpacs,aps,prd,floatfix]{revtex4}

\usepackage{graphicx}
\usepackage{dcolumn}
\usepackage{epsfig}
\usepackage{psfrag}
\usepackage{amsmath}
\newcommand{\BaBarYear}    {08}
\newcommand{\BaBarNumber}  {041}

\newcommand{\SLACPubNumber} {13380}
\newcommand{\LANLNumber} {0809.1174 [hep-ex]}

 \newcommand{\BaBarType}      {PUB}  % Journal publication

\input pubboard/babarsym   % the official way

\RequirePackage{xspace}

\newcommand{\pvec}{{\bf p}}
\newcommand{\Sf}{\ensuremath{S_f}}
\newcommand{\Cf}{\ensuremath{C_f}}

\newcommand{\DE}{\ensuremath{\Delta E}}
\newcommand{\mb}{\ensuremath{m_{\rm ES}}}

\newcommand{\xf}{\ensuremath{{\cal F}}}
\newcommand{\hel}{\ensuremath{{\cal H}}}
\newcommand{\thetaT}{\ensuremath{\theta_{\rm T}}}
\newcommand{\costhr}{\ensuremath{\cos\thetaT}}

\providecommand{\dt}{\deltat}
\newcommand{\ttag}{\ensuremath{t_{\rm tag}}}
\newcommand{\bflav}{\ensuremath{B_{\rm flav}}}

\newcommand\etal{{\it et al.}}
\newcommand{\half}{\ensuremath{\frac{1}{2}}}

\newcommand{\msp}{\ensuremath{\phantom{-}}}

\newcommand{\bfig}{\begin{figure}[htbpc!]}
\newcommand{\efig}{\end{figure}}
\newcommand\bef{\begin{figure}}
\newcommand\edf{\end{figure}}
\newcommand\dbline{\noalign{\vskip 0.10truecm\hrule}\noalign{\vskip 2pt}\noalign{\hrule\vskip 0.10truecm}}

\newcommand\sgline{\noalign{\vskip 0.10truecm\hrule\vskip 0.10truecm}}
\newcommand\beq{\begin{equation}}
\newcommand\eeq{\end{equation}}
\newcommand\bear{\begin{array}}
\newcommand\enar{\end{array}}
\newcommand\beqa{\begin{eqnarray}}
\newcommand\eeqa{\end{eqnarray}}
\newcommand\ben{\begin{enumerate}}
\newcommand\een{\end{enumerate}}

\newcommand{\UfourS}{\ensuremath{\Upsilon(4S)}}

\newcommand{\etagg}{\ensuremath{\eta_{\gaga}}}
\newcommand{\etappp}{\ensuremath{\eta_{3\pi}}}

\newcommand{\etapepp}{\ensuremath{\etapr_{\eta\pi\pi}}}

\newcommand{\etaprg}{\ensuremath{\etapr_{\rho\gamma}}}

\newcommand{\etapeppgg}{\ensuremath{\etapr_{\eta(\gamma\gamma)\pi\pi}}}
\newcommand{\etapeppppp}{\ensuremath{\etapr_{\eta(3\pi)\pi\pi}}}

   \newcommand{\rhoz}{\ensuremath{\rho^0}}

   \newcommand{\fetaprgKp}{\ensuremath{\etapr_{\rho\gamma} K^+}}
   \newcommand{\etaprgKp}{\ensuremath{\Bp\ra\fetaprgKp}}

\newcommand{\fetapKz}{\ensuremath{\etapr K^0}}
\newcommand{\fetapKs}{\ensuremath{\etapr\KS}}

\newcommand{\fetapKl}{\ensuremath{\etapr\KL}}
\newcommand{\etapKz}{\ensuremath{\Bz\ra\fetapKz}}
\newcommand{\etapKs}{\ensuremath{\Bz\ra\fetapKs}}

\newcommand{\etapKl}{\ensuremath{\Bz\ra\fetapKl}}

\newcommand{\sksetap}{\ensuremath{S_{\fetapKs}}}
\newcommand{\cksetap}{\ensuremath{C_{\fetapKs}}}
\newcommand{\skletap}{\ensuremath{S_{\fetapKl}}}
\newcommand{\ckletap}{\ensuremath{C_{\fetapKl}}}
\newcommand{\skzetap}{\ensuremath{S_{\fetapKz}}}
\newcommand{\ckzetap}{\ensuremath{C_{\fetapKz}}}

\newcommand{\SetapKz}{\ensuremath{0.xx\pm0.xx\pm0.xx}}
\newcommand{\CetapKz}{\ensuremath{0.xx\pm0.xx\pm0.xx}}
\newcommand{\fomegaKz}{\ensuremath{\omega K^0}}
\newcommand{\fomegaKs}{\ensuremath{\omega\KS}}

\newcommand{\omegaKs}{\ensuremath{\Bz\ra\fomegaKs}}

\newcommand{\sksomega}{\ensuremath{S_{\fomegaKs}}}
\newcommand{\cksomega}{\ensuremath{C_{\fomegaKs}}}

\newcommand{\SomegaKs}{\ensuremath{0.xx^{+0.xx}_{-0.xx}\pm zz}}
\newcommand{\ComegaKs}{\ensuremath{0.xx^{+0.xx}_{-0.xx}\pm zz}}

\newcommand{\fpizKs}{\ensuremath{\piz\KS}}

\newcommand{\pizKs}{\ensuremath{\Bz\ra\fpizKs}}

\newcommand{\skspiz}{\ensuremath{S_{\fpizKs}}}
\newcommand{\ckspiz}{\ensuremath{C_{\fpizKs}}}

\newcommand{\SpizKs}{\ensuremath{0.xx^{+0.xx}_{-0.xx}\pm zz}}
\newcommand{\CpizKs}{\ensuremath{0.xx^{+0.xx}_{-0.xx}\pm zz}}

\def\stwob{\ensuremath{\sin\! 2 \beta   }\xspace}

\newcommand{\tcp}{\ensuremath{t_{\CP}}}

\renewcommand{\SetapKz}{\ensuremath{0.57\pm0.08\pm0.02}}
\renewcommand{\CetapKz}{\ensuremath{-0.08\pm0.06\pm0.02}}

\renewcommand{\SpizKs}{\ensuremath{0.55\pm 0.20 \pm 0.03}}
\renewcommand{\CpizKs}{\ensuremath{0.13\pm 0.13 \pm 0.03}}

\renewcommand{\SomegaKs}{\ensuremath{0.55^{+0.26}_{-0.29}\pm 0.02}}
\renewcommand{\ComegaKs}{\ensuremath{-0.52^{+0.22}_{-0.20}\pm 0.03}}

\begin{document}

\begin{flushleft}
\babar-\BaBarType-\BaBarYear/\BaBarNumber \\
SLAC-PUB-\SLACPubNumber \\
arXiv:\LANLNumber
\end{flushleft}

%\begin{flushleft}
%    \rm \babar$\;$Analysis Document \#2027 \\
%    Version 7 \\
%    \today \\[.7in]
%\end{flushleft}

\title{\boldmath Measurement of time dependent \CP\ asymmetry parameters 
in \Bz\ meson decays to \fomegaKs, \fetapKz, and \fpizKs
} 

%% author list as of 02-Jul-2008 (523 authors)
%
\author{B.~Aubert}
\author{M.~Bona}
\author{Y.~Karyotakis}
\author{J.~P.~Lees}
\author{V.~Poireau}
\author{E.~Prencipe}
\author{X.~Prudent}
\author{V.~Tisserand}
\affiliation{Laboratoire de Physique des Particules, IN2P3/CNRS et Universit\'e de Savoie, F-74941 Annecy-Le-Vieux, France }
\author{J.~Garra~Tico}
\author{E.~Grauges}
\affiliation{Universitat de Barcelona, Facultat de Fisica, Departament ECM, E-08028 Barcelona, Spain }
% ...
\collaboration{The \babar\ Collaboration}
\noaffiliation

\begin{abstract}
We present measurements of the 
time-dependent \CP-violation parameters $S$ and $C$ in the decays
$\B^0\rightarrow\omega K^0_{\scriptscriptstyle S}$, 
$\B^0\rightarrow\eta^\prime K^0$,
reconstructed as $\eta^\prime K^0_{\scriptscriptstyle S}$ 
and $\eta^\prime K^0_{\scriptscriptstyle L}$,
and $\B^0\rightarrow\pi^0 K^0_{\scriptscriptstyle S}$.
The data sample corresponds to the full 
\mbox{\slshape B\kern-0.1em{\smaller A}\kern-0.1em B\kern-0.1em{\smaller A\kern-0.2em R}}
dataset of $467\times10^6$
$B\kern 0.18em\overline{\kern -0.18em B}{}$ 
pairs produced at the PEP-II asymmetric-energy $e^+e^-$ collider 
at the Stanford Linear Accelerator Center.  The results are  
$S_{\omega K^0_{\scriptscriptstyle S}} = 0.55^{+0.26}_{-0.29}\pm 0.02$, 
$C_{\omega K^0_{\scriptscriptstyle S}} = -0.52^{+0.22}_{-0.20}\pm 0.03$, 
$S_{\eta^\prime K^0} = 0.57\pm0.08\pm0.02$, 
$C_{\eta^\prime K^0} = -0.08\pm0.06\pm0.02$, 
$S_{\pi^0 K^0_{\scriptscriptstyle S}} = 0.55\pm 0.20 \pm 0.03$, and 
$C_{\pi^0 K^0_{\scriptscriptstyle S}} = 0.13\pm 0.13 \pm 0.03$, 
where the first errors are statistical and the second systematic.  These
results are consistent with our previous measurements and the world average
of $\sin\! 2 \beta$ 
measured in 
$B^0\rightarrow J/\psi K^0_{\scriptscriptstyle S}$.

%We present measurements of the 
%time-dependent \CP-violation parameters \skz\ and \ckz\ in the decays
%\omegaKs, \etapKz, reconstructed as \fetapKs\ and \fetapKl, and \pizKs.
%The data sample corresponds to the full \babar\ dataset of $467\times10^6$
%\BB\ pairs produced at the PEP-II asymmetric-energy \epem\ collider 
%at the Stanford Linear Accelerator Center.  The results are  
%$\sksomega = \SomegaKs$, $\cksomega = \ComegaKs$, 
%$\skzetap = \SetapKz$, $\ckzetap = \CetapKz$, 
%$\skspiz = \SpizKs$, and $\ckspiz = \CpizKs$, 
%where the first errors are statistical and the second systematic.  These
%results are consistent with our previous measurements and the world average
%of \stwob\ measured in $\Bz\ra J/\psi\KS$.
\end{abstract}

% PACS, the Physics and Astronomy Classification Scheme.

\maketitle

\section{Introduction}
\label{sec:intro}

Measurements of time-dependent \CP\ asymmetries in \Bz\ meson decays
through $b\rightarrow c \bar{c} s$ amplitudes have provided
crucial tests of the mechanism of \CP\ violation in the Standard Model (SM)
\cite{CPVobsInB}.  These amplitudes contain the leading $b$-quark
couplings,  given by the
Cabibbo-Kobayashi-Maskawa \cite{CKM} (CKM) flavor 
mixing matrix, for kinematically allowed transitions.  
Decays to
charmless final states such as $\phi\Kz$, $\piz\Kz$, 
$\etapr\Kz$, $\omega\Kz$, $\Kp\Km\Kz$, $f_0(980)\Kz$ are CKM-suppressed
 $b\to \qqbar s$ ($q=u,d,s$) processes dominated by a single loop (penguin)
amplitude.  This amplitude has the same weak phase $\beta =
\arg{(-V_{cd} V^*_{cb}/ V_{td} V^*_{tb})}$ of the CKM mixing matrix as that
measured in the $b \to c \bar{c} s$ transition, but is sensitive to
the possible presence of new heavy particles in the loop
\cite{Penguin}.  Due to the different non-perturbative strong-interaction
properties of the various penguin decays,  the effect of new physics is
expected to be channel dependent. 

The CKM phase $\beta$ is accessible experimentally through
interference between the direct decay of the $B$ meson to a 
\CP\ eigenstate and \BzBzb\ mixing followed by decay to
the same final state.  This interference is
observable through the time evolution of the decay.  In the present
study, we reconstruct one \Bz\ from $\FourS\ra\BzBzb$, which decays to
the \CP\ eigenstate \fomegaKs, \fetapKs, \fetapKl, or \fpizKs\
($B_{\CP}$).  
From the remaining particles in the
event we also reconstruct the decay vertex of the other $B$ meson ($B_{\rm
  tag}$) and identify its flavor.  The distribution of the difference
$\deltat \equiv \tcp - \ttag$ of the proper decay times $\tcp$ and $\ttag$
of these mesons is given by
\begin{eqnarray}
 f(\dt) &=& 
 \frac{e^{-\left|\deltat\right|/\tau}}{4\tau} \{1 \pm 
                                                   \label{eq:FCPdefPure}\\
 &&\hspace{-1em}
\left[-\eta_f \Sf\sin(\deltamd\deltat) - \Cf\cos(\deltamd\deltat)\right]\}\,\nonumber
\end{eqnarray}
where $\eta_f$ is the \CP\ eigenvalue of final state $f$ ($-1$ for
\fomegaKs, \fetapKs, and \fpizKs; $+1$ for \fetapKl).  The upper
(lower) sign denotes a decay accompanied by a \Bz (\Bzb) tag, $\tau$
is the mean $\Bz$ lifetime, and $\deltamd$ is the mixing frequency.

A nonzero value of the parameter \Cf\ would indicate direct \CP\ violation.
In these modes we expect $\Cf=0$ and $-\eta_f\Sf= \stwob$, assuming penguin
dominance of the $b \to s$ transition and neglecting other CKM-suppressed
amplitudes with a different weak phase.  However, these CKM-suppressed
amplitudes and the color-suppressed tree diagram introduce additional weak
phases whose contributions may not be
negligible~\cite{Gross,Gronau,BN,london}.  As a consequence, the measured
$S_f$ may differ from \stwob even within the SM.  This deviation $\Delta
S_f=S_f - \stwob$ is estimated in several theoretical approaches: QCD
factorization (QCDF)~\cite{BN,beneke}, QCDF with modeled
rescattering~\cite{Cheng}, soft collinear effective theory
(SCET)~\cite{Zupan}, and SU(3) symmetry~\cite{Gross,Gronau,Jonat}. The
estimates are channel dependent.  Estimates of $\Delta S$ from QCDF are in
the ranges $(0.0,0.2)$, $(-0.03,0.03)$, and $(0.01,0.12)$ for \fomegaKs,
\fetapKz, and \fpizKs, respectively \cite{beneke, Zupan, CCS}; SU(3)
symmetry provides bounds of $(-0.05,0.09)$ for \fetapKz\ and $(-0.06,0.12)$
for \fpizKs\ \cite{Jonat}.  Predictions that use isospin symmetry to
relate several amplitudes, including the $I=\frac{3}{2}$ $B\ra K\pi$ amplitude, give
an expected value for $\skspiz$ near $1.0$ instead of $\stwob$~\cite{Spiks}.

We present updated measurements of mixing-induced \CP\
violation in the \Bz\ decay modes \fomegaKs, \fetapKz, and \fpizKs, which
supersede our previous
measurements~\cite{PreviousOmK,PreviousEtapK,PreviousPizK}.  Significant
changes to previous analyses include twice as much data for \fomegaKs,
$20\%$ more data for \fetapKz\ and \fpizKs, improved track
reconstruction, and an additional decay channel in \fetapKl.  Despite the
modest increase in data, the uncertainties on $\skzetap$ and $\ckzetap$
decrease by $20\%$ and $25\%$, respectively.  Measurements in these modes have
also been made by the Belle Collaboration~\cite{BELLEetapK,BELLEKsPi0}.

\section{The BaBar detector and dataset}

The results presented in this paper are based on data collected with the
\babar\ detector at the \pep2\ asymmetric-energy \epem\ storage ring,
operating at the Stanford Linear Accelerator Center. At \pep2, 9.0 \gev\
electrons collide with 3.1 \gev\ positrons to yield a center-of-mass energy
of $\sqrt{s}=10.58$ \gev, which corresponds to the mass of the \FourS\
resonance.  The asymmetric energies result in a boost from the laboratory to
the \epem\ center-of-mass (CM) frame of $\beta\gamma\approx 0.56$.  We
analyze the entire \babar\ dataset collected at the \FourS\ resonance,
corresponding to an integrated luminosity of 426~fb$^{-1}$ and $(467 \pm
5)\times 10^6$ \BB\ pairs.  We use an additional 44~fb$^{-1}$ of data
recorded about 40 \mev\ below this energy (off-peak) for the study of the
non-\BzBzb\ background.

A detailed description of the \babar\ detector can be found
elsewhere~\cite{BABARNIM}.  Surrounding the interaction point is a
five-layer double-sided silicon vertex tracker (SVT) that 
provides precision measurements near the collision point of
charged particle tracks in the planes transverse
to and along the beam direction. A 40-layer drift chamber (DCH) surrounds
the SVT.  Both of these tracking devices operate in the 1.5~T magnetic
field of a superconducting solenoid to provide measurements of the
momenta of charged particles. 
Charged hadron identification is achieved through
measurements of particle energy loss in the tracking system and the
Cherenkov angle obtained from a detector of internally reflected Cherenkov
light (DIRC). A CsI(Tl) electromagnetic calorimeter (EMC) provides photon
detection, electron identification, and $\piz$, $\eta$, and \KL\
reconstruction. Finally, the instrumented flux return (IFR) of the magnet
allows discrimination of muons from pions and detection of \KL\ mesons.  For
the first 214$\invfb$ of data, the IFR was composed of a resistive plate
chamber system.  For the most recent 212$\invfb$ of data, a portion
of the resistive plate chamber system has been replaced by limited streamer
tubes~\cite{lsta}.

\section{Vertex reconstruction}
\label{sec:Vertexing}

In the reconstruction of the $B_{C\!P}$ vertex, we use all charged daughter tracks.
Daughter tracks that form a \KS\ are fit to a separate vertex, with the
resulting parent momentum and position used in the fit to the $B_{C\!P}$ vertex.
The vertex for the $B_{\rm tag}$ decay is constructed from all tracks in the event 
except the daughters of $B_{C\!P}$. An additional constraint is provided by the 
calculated $B_{\rm tag}$ production point and three-momentum, with its associated 
error matrix. This is determined from the knowledge of the three momentum
of the fully reconstructed $B_{C\!P}$ candidate, its decay vertex and error
 matrix, and from the knowledge of the average position of the \epem\ interaction 
point and \FourS\ average boost.
In order to reduce bias and tails due to long-lived particles, \KS\ and
 $\Lambda^0$ candidates are used as input to the fit in place of their 
daughters. In addition, tracks consistent with photon conversions 
($\gamma\to\epem$) are excluded from the fit.
To reduce contributions from charm decay products that bias 
the determination of the vertex position the tracks with a 
vertex $\chi^2$ contribution greater than 6 are removed and the fit is
repeated until no track fails the $\chi^2$ requirement.  We obtain \deltat\
from the measured distance \deltaz\ between the $B_{C\!P}$ and  $B_{\rm
  tag}$ vertex with the relation  $\Delta z \simeq \beta\gamma c \deltat$.

Because there are no charged particles present at the \pizKs\ decay
vertex, the $\fpizKs$ vertex reconstruction differs significantly from 
that of the \fomegaKs\ and \fetapKz\ analyses. 
In \fpizKs\ we identify the vertex of the $B_{C\!P}$ using the single $\KS$
trajectory from the $\pip\pim$ momenta and the knowledge of the average
interaction point (IP)~\cite{IP}, which is determined several times per hour
from the spatial distribution of vertices from two track events.  
The average transverse size of the IP is $180\ \mu m \times 4\ \mu m$.
We compute
\dt\ and its uncertainty with a geometric fit to the $\FourS \to \BzBzb$
system that takes this IP constraint into account.
We further improve the accuracy of the $\dt$ measurement by constraining the sum of the two
$B$ decay times ($t_{C\!P} +t_{\rm tag}$) to be equal to $2\tau$ ($\tau$ is the
mean \Bz\ lifetime) with an uncertainty $\sqrt{2}\tau$, which effectively
improves the determination of the decay position of the $\FourS$. 
We have verified in a full detector simulation that this procedure
provides an unbiased estimate of \dt.

The estimate of the uncertainty on \dt\ for each \fpizKs\ event reflects the
strong dependence of the \dt\ resolution on the \KS\ flight direction and
on the number of SVT layers traversed by the \KS\ decay daughters.  
When both pion tracks are reconstructed with information from at least the first
three layers of the SVT in the coordinate along the collision axis (axial) as well
as on the transverse plane (azimuthal), 
we obtain \dt\ with resolution comparable 
to that of the \fomegaKs\ and \fetapKz\ analyses.  The average \dt\
resolution in these  modes is about 1.0~ps. 
Events for which there is axial and azimuthal information from the first
three layers of the SVT and for which $\dt$ and the error on $\dt$ satisfy
$|\dt|<20$~ps and $\sigma_{\Delta t}<2.5$~ps are classified as ``good''
(class $g$), and their \dt\ information is used in the time dependent part of
the likelihood function (Eq.~\ref{eq:genPDFg}). About 60\% of the events
fall in this class. Otherwise events are classified as ``bad'' (class $b$).
Since $C_f$ can also be extracted from flavor tagging information alone,
events of class $b$ contribute to the measurement of $C_f$
(Eq.~\ref{eq:genPDFb}) and to the signal yield in the \fpizKs\ analysis.

In \fomegaKs\ and \fetapKz\ decays, the determination of the $B$ decay
vertex is dominated by the charged daughters of the $\omega$ and $\etapr$, so we
do not require information in the first three SVT layers from \KS\ daughter
pions for events in class $g$.  Also, since about 95\% of
events in these modes are of class $g$, the precision of the measurement
of $C_f$ is not improved by including class $b$ events.  We maintain
simplicity of these analyses by simply rejecting class $b$ events.

\section{\boldmath Flavor tagging and \dt\ resolution}
\label{sec:Tagging}

\def\leptontag{{\tt Lepton}}
\def\kaonitag{{\tt Kaon\,I}}
\def\kaoniitag{{\tt Kaon\,II}}
\def\kpitag{{\tt Kaon-Pion}}
\def\piontag{{\tt Pion}}
\def\othertag{{\tt Other}}
\def\notag{{\tt Untagged}}

In the measurement of time-dependent \CP\ asymmetries, it is important
to determine whether at the time of decay of the $B_{\rm tag}$
the $B_{CP}$ was a \Bz\ or a \Bzb. This `flavor tagging'
is achieved with the analysis of the decay products of the recoiling 
$B_{\rm tag}$ meson.
Most \B\ mesons decay to a final state that is 
flavor specific; i.e., only accessible from either a \Bz\ or a \Bzb, 
but not from both. The purpose of the flavor tagging 
algorithm is to determine the flavor of $B_{\rm tag}$ with the highest possible
efficiency $\epsilon$ and lowest possible probability \mistag\ 
of assigning a wrong flavor to $B_{\rm tag}$. 
The figure of merit for the performance of the tagging algorithm is the 
effective tagging efficiency 
\begin{equation}
Q = \epsilon (1-2\mistag)^2,
\end{equation}
which is approximately related to the statistical uncertainty $\sigma$ in the 
coefficients $S$ and $C$ through
\begin{equation}
\sigma \propto \frac{1}{\sqrt{Q}}.
\end{equation}
It is not necessary to reconstruct $B_{\rm tag}$ fully to determine its flavor.
We use a neural network based technique~\cite{babarsin2betaprd} to exploit signatures of $B$ decays that
determine the flavor at decay of the $B_{\rm tag}$.
Primary leptons from semileptonic $B$ decays are
selected from identified electrons and muons as well as isolated
energetic tracks.  The charges of identified kaon candidates define a
kaon tag.  Soft pions from \Dstarp decays are selected on the basis of
their momentum and direction with respect to the thrust axis of $B_{\rm
tag}$.  Based on the output of this algorithm, candidates are divided into
seven mutually exclusive categories.  These are (in order of decreasing
signal purity) \leptontag, \kaonitag, \kaoniitag, \kpitag, \piontag,
\othertag, and \notag.   

We apply this algorithm to a sample of fully reconstructed, self-tagging,
neutral \B decays ($B_{\rm flav}$ sample). We use \B decays to
$D^{(*)-}(\pi^+,\rho^+, a_1^+)$ to measure the tagging efficiency $\epsilon$,
mistag rate $w$, and the difference in mistag rates for \Bz\ and \Bzb\ tag-side 
decays $\Delta w \equiv w(\Bz)-w(\Bzb)$. The results are shown
in Table~\ref{tab:mistag}. The \notag\ category of events contains no flavor 
information and therefore carries no weight in the time-dependent analysis.  
The total effective tagging efficiency $Q$ for this algorithm is measured to be
$(31.2 \pm 0.3)\%$.    

\begin{table}[!t]
\caption
{Efficiencies $\varepsilon$, average mistag fractions $\mistag$, mistag fraction differences
$\Delta\mistag \equiv \mistag(\Bz)-\mistag(\Bzb)$, and effective tagging
  efficiency $Q \equiv \epsilon (1-2\mistag)^2$ for each tagging category from
  the $B_{\rm flav}$  data.
}
\label{tab:mistag} 
\begin{ruledtabular} 
\begin{tabular*}{\hsize}{l
@{\extracolsep{0ptplus1fil}}  D{,}{\ \pm\ }{-1} 
@{\extracolsep{0ptplus1fil}} D{,}{\ \pm\ }{-1} 
@{\extracolsep{0ptplus1fil}}  D{,}{\ \pm\ }{-1} 
@{\extracolsep{0ptplus1fil}}  D{,}{\ \pm\ }{-1}}  
Category     & 
\multicolumn{1}{c}{$\ \ \ \varepsilon$   (\%)} & 
\multicolumn{1}{c}{$\ \ \ \mistag$       (\%)} & 
\multicolumn{1}{c}{$\ \ \ \Delta\mistag$ (\%)} &
\multicolumn{1}{c}{$\ \ \ Q$             (\%)} \\ \colrule  
  \leptontag &   9.0,0.1&   2.8,0.3 &  0.3,0.5&    8.0,0.1 \\
   \kaonitag &  10.8,0.1&   5.3,0.3 &  -0.1,0.6&    8.7,0.1 \\
  \kaoniitag &  17.2,0.1&  14.5,0.3 &  0.4,0.6&    8.7,0.2 \\
     \kpitag &  13.7,0.1&  23.3,0.4 &  -0.7,0.7&    3.9,0.1 \\
    \piontag &  14.2,0.1&  32.5,0.4 &   5.1,0.7&    1.7,0.1 \\
   \othertag &  9.5,0.1&  41.5,0.5 &   3.8,0.8&    0.3,0.0 \\
\colrule
         All &  74.4,0.1&           &          &   31.2,0.3 \\
\end{tabular*} 
\end{ruledtabular} 
\end{table}

Including the effects of the mistag rate, Eq.~\ref{eq:FCPdefPure} becomes
\begin{eqnarray}
  F(\dt) &=&
        \frac{e^{-\left|\deltat\right|/\tau}}{4\tau} \{1 \mp\Delta w \pm
                                                   \label{eq:FCPdef}\\
   &&\hspace{-2em}(1-2w)
\left[-\eta_f \Sf\sin(\deltamd\deltat) - \Cf\cos(\deltamd\deltat)\right]\}.\,\nonumber
\end{eqnarray}
Finally, to account for experimental $\Delta t$ resolution, we convolve
Eq.~\ref{eq:FCPdef} with a resolution function, the parameters of which we
obtain from fits to the \bflav\ sample.  The \dt\ resolution
function is represented as a sum of three Gaussian distributions with
different widths.  For the core and tail Gaussians, the widths are
scaled by $\sigma_{\Delta t}$.  
In addition we allow an offset for the core distribution in the hadronic
tagging categories (Sec.~\ref{sec:Tagging}) separate from that of
the \leptontag\ category, to allow for a small bias of \dt\ from
secondary $D$-meson decays; a common offset is used for the tail component.
The third Gaussian (of fixed 8 ps width) accounts for the few events with
incorrectly reconstructed vertices.   Identical resolution
function parameters are used for all $B_{\CP}$ modes, since the $B_{\rm tag}$ 
vertex precision dominates the \dt resolution.   

Events without reliable \deltat information (class $b$) are
sensitive to the parameter $\Cf$ and are used to constrain this
parameter in the \fpizKs\ analysis.  Integrating Eq.~\ref{eq:FCPdef}
over \deltat we get
\begin{equation}
F^C  =\frac{1}{2}\;  \left\{1 \mp \left[\Delta w + \Cf
(1-2w)/(1+\Delta m_d^2 \tau^2)\right]\right\}. \label{eq:FCbaddt}
\end{equation}
We also account for the asymmetry in tagging efficiency for \Bz and \Bzb
decays, but, for simplicity, we assume the asymmetry is zero in the
above equations.

\section{Event reconstruction and selection}

We choose event selection criteria with the aid of a detailed
Monte Carlo (MC) simulation of the \B\ production and decay sequences,
and of the detector response \cite{geant}.  These criteria are designed
to retain signal events with high efficiency while removing most of the
background.

We reconstruct the $B_{CP}$ candidate by combining the four-momenta of
the two daughter mesons, with a vertex constraint.
The \B-daughter candidates are reconstructed with the following decays:
$\piz\ra\gaga$; $\eta\ra\gaga$ (\etagg); $\eta\ra\pip\pim\piz$
(\etappp); $\etapr\ra\etagg\pip\pim$ (\etapeppgg);
$\etapr\ra\etappp\pip\pim$ (\etapeppppp); $\etapr\ra\rhoz\gamma$
(\etaprg), where $\rhoz\ra\pip\pim$; $\omega\ra\pip\pim\piz$; and
$\KS\ra\pip\pim$ 
($K^0_{\pi^+\pi^-}$).  In the \fetapKs\ analysis we also reconstruct $\KS$
via its decay to two neutral pions ($K^0_{\pi^0\pi^0}$).  
The requirements on the invariant
masses of these particle combinations are given in Table \ref{tab:rescuts}.    
We consider as photons energy depositions in the EMC that are isolated
from any charged tracks, carry a minimum energy of $30\mev$, and have
the expected lateral shower shapes. 

The five final states used for \etapKs\ are 
\etapeppgg$K^0_{\pi^+\pi^-}$,
\etaprg$K^0_{\pi^+\pi^-}$,
\etapeppppp$K^0_{\pi^+\pi^-}$,
\etapeppgg$K^0_{\pi^0\pi^0}$,
and \etaprg$K^0_{\pi^0\pi^0}$.
For the \etapKl\ channel we reconstruct the \etapr\ in two modes:
\etapeppgg\ and
\etapeppppp.
Large backgrounds to the final states $\etaprg\KL$, $\omega\KL$, and
$\omega\Kz_{\piz\piz}$ preclude 
these modes from improving the precision of the measurement of \CP\
parameters in \fetapKz\ and \fomegaKz; $\pi^0\KL$ and
$\piz\Kz_{\piz\piz}$ events lack the minimum information for
reconstruction of the decay vertex.   

For decays with a $\KS\ra\pi^+\pi^-$ candidate we perform a fit of the
entire decay tree which constrains the \KS\ flight 
direction to the pion pair momentum direction and the \KS\
production point to the $B_{C\!P}$ vertex determined as
described in Sec.~\ref{sec:Vertexing}.  In this vertex fit we also
constrain the $\eta$, \etapr, and \piz\ candidate masses to 
world-average values \cite{PDG2006}, since these resonances have natural
widths that are negligible compared to the resolution. Given that the
natural widths of the $\omega$ and $\rho$ mesons are comparable to or greater than the
detector resolution, we do not impose any constraint on the masses of these
candidates; constraining the mass of the \KS\ does not
improve determination of the vertex.  In the \fomegaKs\ and \fetapKs\
analyses, we require the $\chi^2$ probability of this fit to be greater than
0.001.  We also require that the \KS\ flight length divided by its
uncertainty be greater than $3.0$ ($5.0$ for \pizKs).

\begin{table}[!hbtp]
\begin{center}
\caption{
Selection requirements on the invariant masses of candidate resonant 
decays and the laboratory energies of photons from the decay.}
\label{tab:rescuts}
\begin{tabular}{lcc}
\dbline
State		& Invariant mass (MeV)		& $E(\gamma)$ (MeV)\\
\sgline						
Prompt \piz	& $110 < m(\gamma\gamma) < 160$		& $>50$		\\
Secondary \piz	& $120 < m(\gamma\gamma) < 150$		& $>30$		\\
\etagg		& $490 < m(\gamma\gamma) < 600$		& $>50$	\\
\etappp		& $520 < m(\pip\pim\piz) < 570$		& ---		\\
\etapepp	& $945 < m(\pip\pim\eta) <970$		& ---		\\
\etaprg		& $930 < m(\pip\pim\gamma) <980$	& $>100$	\\
$\omega$	& $735 < m(\pip\pim\piz)  < 825$& ---	\\
\rhoz		& $470 < m(\pip\pim) <980$		& ---		\\
\vspace{0.3mm}
$\Kz_{\pip\pim}$& $486 < m(\pip\pim) <510$		& ---		\\
$\Kz_{\piz\piz}$& $468 < m(\piz\piz) <528$		& ---		\\
\dbline
\end{tabular}
\vspace{-5mm}
\end{center}
\end{table}

Signal \KL\ candidates are reconstructed from clusters of energy deposited in
the EMC or from hits in the IFR not associated with any charged track in
the event \cite{Stefan}.  Taking the flight direction of the \KL\ to be the
direction from the \Bz decay vertex to the cluster centroid, we determine
the $\KL$ momentum $\pvec_{\KL}$ from a fit with the $B^0$ and \KL\ masses
constrained to world-average values ~\cite{PDG2006}.

\subsection{\boldmath Kinematics of $\UfourS\ra\BB$}

In this experiment the energy of the initial \epem\ state is
known within an uncertainty of a few \mev. For a final state with two particles we can
determine four kinematic variables from conservation of energy and
momentum.  These may be taken as polar and azimuthal angles of the
line of flight of the two particles, and two energy, momentum, or mass
variables, such as the masses of the two particles.  In practice, since
we fully reconstruct one \B\ meson candidate, we make the assumption
that it is one of two final-state particles of equal mass.  We compute
two largely uncorrelated variables that test consistency with this
assumption, and with the known value \cite{PDG2006} of the \B-meson
mass.  The choice of these variables depends on the decay process, as
we discuss below. 

In the reconstruction of \omegaKs\ and \etapKs\ the kinematic variables
are the energy-substituted mass
\begin{eqnarray}
\mes &\equiv& \sqrt{(\half s + \pvec_0\cdot\pvec_B)^2/E_0^2 - \pvec_B^2}
\end{eqnarray}
and the energy difference
\begin{eqnarray}
\DE &\equiv& E_B^*-\half\sqrt{s}
\end{eqnarray}
where $(E_0,\pvec_0)$ and $(E_B,\pvec_B)$ are the laboratory four-momenta of
the \UfourS\ and the $B_{C\!P}$ candidate, respectively, and the asterisk
denotes the \UfourS\ rest frame.  The resolution is $3\ \mev$ in \mes\ and
$20$-$50\ \mev$ in \DE, depending on the decay mode.  We require
$5.25<\mes<5.29\ \gev$ and $|\DE|<0.2$ \gev, as distributions of these
quantities for signal events peak at the $B$-meson mass in \mes\ and zero
in \DE.  

For the \etapKl\ channel only the direction of the \KL\ momentum is
measured.  For these  candidates \mes\ is not determined; instead
we obtain \DE\ from a calculation with the $B^0$ and
\KL\ masses constrained to world-average values.  Because of the mass
constraint on the  $B^0$, the \DE\ distribution for \KL events, which 
peaks at zero for signal, is asymmetric and narrower than that of \KS\
events; we require $-0.01<\DE<0.08$ \gev for \fetapKl.  

\def\mmiss {\ensuremath{m_{\rm miss}}}

For the \fpizKs\ analysis we use the kinematic variables $m_B$ and $m_{\rm miss}$.
The variable $m_B$ is the invariant mass of the reconstructed $B_{C\!P}$. The
variable $m_{\rm miss}$ is the invariant mass of the $B_{\rm tag}$,
computed from the known beam energy and the measured $B_{CP}$
momentum with $m(B_{CP})$ constrained to the nominal $B$-meson mass $m_B^{PDG}$
~\cite{MassConstraint}.  For signal decays, $m_B$ and $m_{\rm miss}$
peak at the $B^0$ mass and have resolutions of $\sim 47 \mev$ and $\sim
5.4 \mev$, respectively; the distribution of $m_B$ exhibits a low-side
tail due to 
leakage of energy out of the EMC.  To compare the \mmiss\ resolution with the \mes\ resolution
a factor of two from the approximate relation $\mes \sim (\mmiss +
m_B^{PDG})/2$ should be taken into account. The beam-energy constraint
in $m_{\rm miss}$ helps to eliminate the correlation
with $m_B$. 
We select candidates within the window
$5.11 < m_{\rm miss} < 5.31$ \gev and $5.13 < m_B < 5.43$ \gev, which
includes the signal peak and a sideband region for background
characterization.

\subsection{Background reduction}
\label{sec:Background}

Background events arise primarily from random combinations of particles
in continuum $\epem\ra\qqbar$ events ($q=u,d,s,c$).  For some of the
decay chains we must also consider cross feed from \B-meson decays by
modes other than the signal; we discuss these in Secs.~\ref{sec:Components}
and \ref{sec:PDFs} below.  To reduce the \qqbar\ backgrounds we make use of
additional properties of the event that are consequences of the 
decay.

For the \omegaKs\ and \etapKz\ channels we define the angle \thetaT\ between the
thrust axis \cite{thrust} of the $B_{C\!P}$ candidate in the \UfourS\ frame
and that of the charged tracks and neutral calorimeter clusters in the
rest of the event.  The event is required to contain at least
one charged track not associated with the $B_{C\!P}$ candidate.  The
distribution of $|\costhr|$ is sharply peaked near 1 for \qqbar\ jet
pairs, and nearly uniform for \B-meson decays.  The requirement
is $|\costhr|<0.9$ for all modes.

For the \etaprg\ decays we also define the angle $\theta_{\rm
dec}^\rho$ between the momenta of the \rhoz\ daughter \pim\ and of the
\etapr, measured in the \rhoz\ rest frame.  We require
$|\cos\theta^{\rho}_{\rm dec}|< 0.9$ to reduce the combinatorial
background of \rhoz\ candidates incorporating a soft pion that are
reconstructed as decays with $|\cos\theta^{\rho}_{\rm dec}|\simeq 1$. 

For \fetapKl\ candidates we require that the cosine of the polar angle of
the total missing momentum in the laboratory system be less than 0.96
to reject very forward \qqbar\ jets. 
We construct the missing momentum $\pvec_{\rm miss}$ as the difference of 
$\pvec_0$ and the momenta of all charged tracks and neutral clusters
not associated with the \KL\ candidate. We project $\pvec_{\rm miss}$
onto $\pvec_{\KL}$,  
and require the component perpendicular to the beam line, $p_{{\rm miss}\perp}^{\rm
proj}$, to satisfy $p_{{\rm miss}\perp}^{\rm proj}-p_{\KL\perp} > -0.8\
\gev$.  These values are chosen to minimize the 
uncertainty on $S$ and $C$ in the presence of background.

The purity of the \KL\ candidates
reconstructed in the EMC is further improved by a requirement on the
output of a neural network (NN) that takes cluster-shape variables as
inputs.  For the NN, we use the following eight variables: the number of crystals in the EMC
cluster; the total energy deposited in the EMC cluster; the second moment of the cluster energy
\begin{eqnarray}
\mu_2&=&\frac{\sum_{i} E_i\cdot r_i^2}{\sum_{i} E_i},
\end{eqnarray} 
where $E_i$ is the energy deposited in the $i^{\rm th}$ crystal and $r_i$ 
its distance from the cluster centroid; 
the lateral moment 
\begin{eqnarray}
\mu_{\rm LAT}&=&\frac{\sum_{i=2,n} E_i \cdot r_i^2}
{(\sum_{i=2,n} E_i \cdot r_i^2) + 25(E_0 + E_1) },
\end{eqnarray}
where $E_0$ refers to 
the most energetic crystal and $E_n$ to the least energetic one; the ratio
$S1/S9$ of the energy of the most energetic crystal ($S1$) to the sum of energy
of the $3\times3$ crystal block with $S1$ in its center ($S9$); the ratio $S9/S25$,
where $S25$ is the sum of energy of the $5\times5$ crystal block with $S1$ in its 
center; and the absolute value of the expansion coefficients
$|Z_{20}|$ and $|Z_{42}|$ of the 
spatial energy distribution 
of the EMC cluster expressed as a series of Zernike polynomials ($\zeta$):
$E(x,y)=\sum_{n,m} Z_{n,m} \cdot  \zeta_{n,m}(r,\phi)$, where
$(x,y)$ are the cartesian coordinates in the plane of the calorimeter,
$(r,\phi)$ are the polar coordinates of the Zernike polynomials
($0 \le r \le 1$) and $n,\,m$ are non-negative integers.  
The NN is trained on MC signal events and off-peak data reconstructed for
the $\etapr_{\eta_{(\gamma\gamma)}\pi\pi}\KL$ decay mode to return +1 if the
event is signal-like and $-1$ if it is background-like.  We check the
performance of the NN on an independent sample of MC signal events and
off-peak data reconstructed for the
$\etapr_{\eta_{(\gamma\gamma)}\pi\pi}\KL$ decay mode.  Using ensembles of
simulated experiments, as discussed in Sec.~\ref{sec:Validation}, we find
that requiring the output of the NN to be greater than $-0.2$ minimizes the
average statisticial uncertainty on $S$ and $C$.

For the \fpizKs\ channel we require the \chisq\ probability of the
kinematic fit to be greater than 0.001.  We exclude
events in which the absolute value of cosine of the angle between the beam
axis and the $B_{\CP}$ momentum in the \UfourS\ frame ($\cos\theta_B$) is
greater than 0.9.  Finally, we apply a cut on the event shape, selecting
events with $L_2 / L_0< 0.55$; $L_i$ is the $i^{\rm th}$ angular
moment defined as $ L_i = \sum_j p_j\times\left|\cos\theta_j\right|^i,$ 
where $\theta_j$ is the angle with
respect to the $B$ thrust axis of track or neutral cluster $j$, $p_j$ is its
momentum, and the sum excludes the daughters of the $B$ candidate.

The average number of candidates found per selected event in \fetapKz\
and \fomegaKs\ is between 1.08 and 1.32, depending on the final
state. For events with multiple candidates we choose the one with the
largest decay vertex probability for the \B\ meson.  Furthermore, in
the \etapKl\ sample, if several $B$ candidates have the same vertex
probability, we choose the candidate with the \KL\ information taken
from, in order, EMC and IFR, EMC only, or IFR only.  From the
simulation we find that this algorithm selects the correct-combination
candidate in about two thirds of the events containing multiple
candidates. 

For the \fpizKs{} channel the average number of candidates found per
selected event is $1.03$. In events with multiple candidates we
choose the one with the smallest value of $\chi^2$ obtained from
the reconstructed mass of the \KS and the \piz candidates and their
respective errors.

In the \fetapKz\ analysis, we estimate from MC the fraction of events in
which we misreconstruct charged daughters of the $\etapr$ (self-crossfeed events),
which dominate the determination of the $B_{CP}$ decay vertex.  We find that
$(1$-$4)\%$ of events are self-crossfeed, depending on the \etapr\ and
$K^0$ decay channel.

\section{Maximum likelihood fit}
\label{sec:MLfit}

The selected sample sizes are given in the first column of
Table~\ref{tab:Results}.  
We perform an unbinned maximum likelihood (ML) fit
to these data to obtain the common \CP-violation parameters and signal
yields for each channel.  
For each signal or background component $j$, tagging category $c$, and event
class $t = g$ (Sec.~ \ref{sec:Vertexing}), we define a total probability
density function (PDF) for event $i$ as 
\begin{equation}
 {\cal P}_{j,c,g}^i \equiv 
 { \cal  T}_{j,c} (\deltat^i, \sigma_{\deltat}^i, \varphi^i) \cdot 
\prod_k {\cal Q}_{k,j} ( x_k^i )\,,   \label{eq:genPDFg}
\end{equation}
where ${ \cal  T}$ is the function $F(\dt)$ defined in
Eq.~\ref{eq:FCPdef} convolved with the \dt\ resolution function, and
$\varphi=\pm1$ is the \B\ flavor defined by upper and lower signs in
Eq.\ref{eq:FCPdefPure}.  For event class $t = b$ (for the \fpizKs\
analysis), we define a total PDF for event $i$ as 
\begin{equation}
 {\cal P}_{j,c,b}^i \equiv 
 F^C_{j,c} (\varphi^i) \cdot 
\prod_k {\cal Q}_{k,j} ( x_k^i )\,,   \label{eq:genPDFb}
\end{equation}
where $F^C$ is the function defined in Eq.~\ref{eq:FCbaddt}.  The set of
variables $x_k^i$, which serve to discriminate signal from background, are
described along with their PDFs $ {\cal Q}_{k,j} (x_k )$ in Sec.~\ref{sec:PDFs} 
below.  The factored form of the PDF is a good approximation
since linear correlations are smaller than 5\%, 7\% and 6\% in the
$\omega\KS$, $\etapr K^0$ and $\piz\KS$ analyses, respectively.  The effects
of these correlations are estimated as described in
Sec.~\ref{sec:Validation}.

We write the extended likelihood function for all candidates for decay
mode $d$ as
\begin{eqnarray}\nonumber
{{\cal L}_d} = \prod_{c,t}
\frac{\exp{\left(-\sum_j{n_jf_{j,t}\epsilon_{j,c}}\right)}}  {N_{c,t}!} \times \\
\prod_i^{N_{c,t}} \left[\sum_{j}n_j f_{j,t}\epsilon_{j,c}{\cal P}_{j,c,t}^i \right]\,,
\end{eqnarray}
where $n_j$ is the yield of events of component $j$, $f_{j,t}$ is the
fraction of events of component $j$ for each event class $t$,
$\epsilon_{j,c}$ is the efficiency of component $j$ for each tagging
category $c$, and $N_{c,t}$ is the number of events of category $c$ with event class
$t$ in the sample.  When combining decay modes we form the grand likelihood
${\cal L}=\prod{\cal L}_d$.   

We fix $\epsilon_{j,c}$ for all components except the \qqbar
background to $\epsilon_{\bflav,c}$, which are listed in
Table~\ref{tab:mistag}. For the \fpizKs\ channel we assume the same $\epsilon_{j,c}$
for class $g$ and class $b$ events.  For \fomegaKs\ and \fetapKz\ we fix
$f_{j,g} = 1$ because we accept only events of class $g$.

\subsection{Model components}
\label{sec:Components}

For all of the decay chains we include in the ML fit a component for
\qqbar\ combinatorial background ($j = \qqbar$), in addition
to the one for the signal. The functional forms of the PDFs that describe this
background are determined from fits of one observable at
a time to sidebands of the data in the kinematic variables that
exclude signal events.  Some of the parameters of this PDF are
free in the final fit.  Thus the combinatorial component receives
contributions from all non-signal events in the data.  

We estimate from the simulation that charmless $B$ decay modes contribute
less than $2\%$ of background to the input sample. These events have
final states different from the signal, but similar kinematics, and
exhibit broad peaks in the signal regions of some observables.  We
find that the charmless \BB\ background component ($j =  \rm{chls}$)
is needed for the final states \fomegaKs, \etaprg$K^0_{\pi^+\pi^-}$,
and \etaprg$K^0_{\pi^0\pi^0}$.  We account for these
with a separate component in the PDF.  Unlike the other fit components, we
fix the charmless \BB\ yields using measured branching fractions, where
available, and detection efficiencies determined from MC.  For unmeasured
background modes, we use theoretical estimates of branching fractions.

We also consider the presence of \B decays to charmed particles in the input sample.
The charmed hadrons in these final states tend to be too heavy to be
misreconstructed as the two light bodies contained in our signals, and their
distributions in the \B\ kinematic variables are similar to those for
\qqbar.  However, in the event shape variables and \dt\ they are
signal-like.  We have found that biases in the fit results are minimized for
the modes with \etaprg\ by including a component specifically for the \B
decays to charm states ($j = \rm{chrm}$).  Finally, for \etapeppppp\KL\ we
divide the signal component into two categories for correctly reconstructed
and self-crossfeed events; we fix the fraction of the self-crossfeed
category to the value obtained from MC.

\subsection{Probability density functions}
\newcommand{\Bztojpsiks}{\ensuremath{\Bz\to\jpsi\KS}}
\newcommand{\Bztokspiz} {\ensuremath{\Bz \to \KS\piz}}
\label{sec:PDFs}

The set of variables $x_k$ of Eqs.~\ref{eq:genPDFg} and \ref{eq:genPDFb}
is defined for each family of decays as 
\begin{itemize}
\item \omegaKs:  $\{\mes, \DE, \xf, m(\pip\pim\piz)\equiv m_\omega, \hel\}$,
\item \etapKs:  $\{\mes, \DE, \xf\}$,
\item \etapKl:  $\{\DE, \xf\}$,
\item \pizKs:  $\{m_B, m_{\rm miss}, L_2/L_0, \cos\theta_B\}$.
\end{itemize}
Here \xf\ is a Fisher discriminant
described below, and \hel\ is the cosine of the polar angle of the normal to
the $\omega$ decay plane in the $\omega$ helicity frame, which is defined as
the $\omega$ rest frame with polar axis opposite to the direction of the $B$.
From Monte Carlo studies we find that including the $\etapr$ mass, $\rho$
mass, $\rho$ helicity, or $\omega$ Dalitz plot coordinates does not improve
the precision of the measurements of $S$ and $C$.

\begin{figure}[!htb]
\hspace*{-0.5cm}
\includegraphics[angle=0,width=0.5\textwidth]{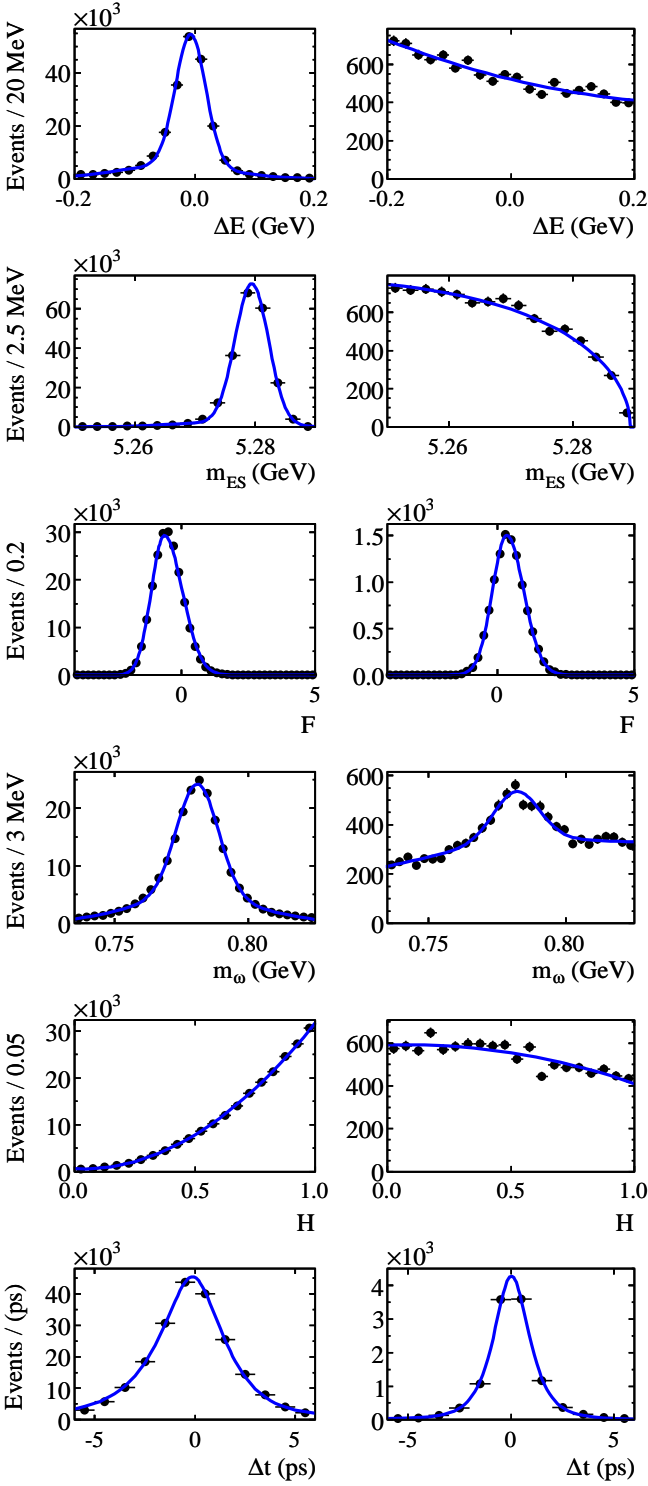}\\
 \caption{\label{fig:pdfs} PDFs for \fomegaKs; from top to bottom \DE, 
 \mes, \xf, $\omega$ mass, $\hel$, and \deltat. In the left column
 we show distributions from signal Monte Carlo; in the right column we show 
 distributions for the \qqbar component, which are taken from sidebands of
 the data in the kinematic variables that exclude signal events. 
} 
\end{figure}

In Fig.~\ref{fig:pdfs} we show PDFs for the signal and \qqbar\
components for the \fomegaKs\ analysis, which are similar to those for
the \fetapKs\ analysis.  We parameterize the PDFs for the signal
component using simulated events, while the background distributions
are taken from sidebands of the data in the kinematic variables that
exclude signal events. The parameters used in the PDFs are different for
each mode. 

For the background PDF shapes we use the following:
%%%%%%%%%%%%%%%%%%%%%%%%%%%%%%%%
% DE, mES, m_B, m_miss
%%%%%%%%%%%%%%%%%%%%%%%%%%%%%%%%%
the sum of two Gaussians for ${\cal Q}_{\rm sig}(\mes)$ and ${\cal Q}_{\rm
sig}(\DE)$;
a quadratic dependence for ${\cal Q}_{\qqbar}(\DE)$, ${\cal Q}_{\rm
chrm}(\DE)$, and ${\cal Q}_{\qqbar}(m_B)$; and the sum of two Gaussians
for ${\cal Q}_{\rm chls}(\DE)$.
% bkg mES, m_miss
For ${\cal Q}_{\qqbar}(\mes)$ and ${\cal Q}_{\qqbar}(m_{\rm miss})$ we use
the function 
\begin{equation}
f(x) = x\sqrt{1-x^2}\exp{\left[-\xi(1-x^2)\right]},
\end{equation}
with $x\equiv2\mes/\sqrt{s}$ ($2\mmiss/\sqrt{s}$ for \fpizKs) and $\xi$ a
free parameter \cite{Argus}, and the same function plus a
Gaussian for ${\cal Q}_{\rm chrm}(\mes)$ and ${\cal Q}_{\rm chls}(\mes)$.
%Cruijff
For ${\cal Q}_{\rm sig}(m_B)$ and ${\cal Q}_{\rm sig}(m_{\rm miss})$ we use the
function
\begin{equation}
f(x) =
\exp \left(\frac{-(x-\mu)^2} {2 \sigma^2_{L,R}+\alpha_{L,R} (x-\mu)^2}
\right),
\end{equation}
where $\mu$ is the peak position of the distribution,
$\sigma_{L,R}$ are the left and right widths, and
$\alpha_{L,R}$ are the left and right tail parameters.
% DE for etapKL
For ${\cal Q}_{\qqbar}(\DE)$ in the \fetapKl\ analysis, we use the function
\begin{equation}
f(x) = x(1-x)^{-2}\exp{\left[\xi^{\prime} x\right]}
\end{equation}
where $x\equiv\DE - (\DE)_\mathrm{min}$, with $(\DE)_\mathrm{min}$ fixed to $-0.01$, and
$\xi^{\prime}$ is a free parameter.

\begin{figure}[!tbp]
\begin{center}
\includegraphics[width=1\linewidth]{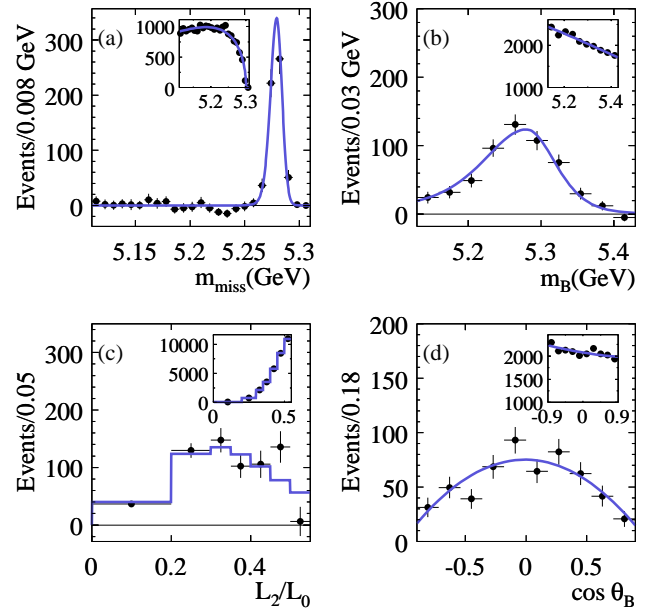}
\caption{Distribution of (a) $\mmiss$, (b) $m_B$, (c) $L_2/L_0$,
(d) $\cos\theta_B$, for signal (background-subtracted) events in data
(points) from the \Bztokspiz\ sample.  
The solid curve represents the shape of the signal PDF, as
obtained from the ML fit. The insets show the distribution of the data,
and the PDF, for background (signal-subtracted) events.}
\label{fig:kspi0splots}
\end{center}
\end{figure}

%%%%%%%%%%%%%%%%%%%%
%Event Shape
%%%%%%%%%%%%%%%%%%%%
To reduce \qqbar\ background beyond that obtained with the \costhr\
requirement described above for \fomegaKs\ and \fetapKz\ (and the
$\theta_B$ and $L_2/L_0$ requirements for \fpizKs), we use additional event topology
information in the ML fit.  The variables used include $\theta_B$, 
$L_{0}$, $L_{2}$, and the angle with respect to the beam axis in the
\UfourS\ frame of the signal $B$ thrust axis ($\theta_S$).
For the \fpizKs\ analysis, we use $\cos\theta_B$ and the ratio
$L_2/L_0$ directly in the fit, parameterized by a second-order
polynomial and a seven-bin histogram, respectively.  
The parameters of the $L_2/L_0$ PDF depend on the tagging category
$c$ in the signal component.  In Fig.~\ref{fig:kspi0splots} we show the PDFs
for signal (background) superimposed on the distribution for data where
background (signal) events are subtracted using an event weighting
technique~\cite{ref:splots}. The bin widths of the $L_2/L_0$ histogram have been
adjusted to be coarser where the background is small to reduce the number of
free parameters of the PDF.

For the other decay modes we construct a Fisher discriminant \xf, which is an 
optimized linear combination of $L_0$, $L_2$, $|\cos{\theta_B}|$, and
$|\cos{\theta_S}|$. For the \KL\ modes we also use the continuous
output of the flavor tagging algorithm as a variable entering the
Fisher discriminant. The coefficients used to combine these
variables are chosen to maximize the separation (difference of means
divided by quadrature sum of errors) between the signal and continuum 
background distributions of \xf , and are determined from studies of signal 
MC and off-peak data.  We have studied the optimization of \xf\ for a variety
of signal modes, and find that a single set of coefficients is
nearly optimal for all.  

The PDF shape for \xf\ is an asymmetric Gaussian with different
widths below and above the peak for signal, plus a broad Gaussian for
\qqbar\ background to account for a small tail in the signal \xf\ region.
The background peak parameter is adjusted to be the same for all tagging 
categories $c$.  Because \xf\ describes the overall shape of the event, 
the distribution for \BB\ background is similar to the signal distribution.

%%%%%%%%%%%%%%%%%%%%%%%%%
% omega mass, hel
%%%%%%%%%%%%%%%%%%%%%%%%
For ${\cal Q}_{\rm sig}(m_\omega)$ we use the sum of two Gaussians; for
${\cal Q}_{\qqbar}(m_\omega)$ and ${\cal Q}_{\rm chls}(m_\omega)$ the
sum of a Gaussian and a quadratic.  For ${\cal Q}_{\rm sig}(\hel)$
and ${\cal Q}_{\qqbar}(\hel)$ we use a quadratic dependence,
and for ${\cal Q}_{\rm chls}(\hel)$ a fourth-order polynomial.
%
%%%%%%%%%%%%%%%%%%%%%%%%%%%%%%%%
% Resolution Model
%%%%%%%%%%%%%%%%%%%%%%%%%%%%%%%%%
%
As described in Sec.~\ref{sec:Tagging}, the resolution function
in ${ \cal T}_j(\deltat)$
is a sum of three Gaussians for all fit components $j$.  For \qqbar\
background we use the same functional form ${ \cal T}_j(\deltat)$ as for
signal, but fix the \B\ lifetime $\tau$ to zero so that ${ \cal
T}_j(\deltat)$ is effectively just the resolution model.
%%%%%%%%%%%%%%%%%%%%%%%%%%%%%%%%
% Floating Fixed Params
%%%%%%%%%%%%%%%%%%%%%%%%%%%%%%%%%
For the signal and \BB\ background components we determine the parameters of
${\cal Q}_{k,j} ( x_k^i )$ from simulation, and the \qqbar\ background
parameters are free in the final fit.  For the signal resolution function 
we fix all parameters to values obtained from the \bflav\ sample; we obtain
parameter values from MC for the charm and charmless \BB\ resolution models;
we leave parameters of the \dt\ resolution model for \qqbar\ free in the final fit.

%%%%%%%%%%%%%%%%%%%%%%%%%%%%%%%%
% Control Samples
%%%%%%%%%%%%%%%%%%%%%%%%%%%%%%%%%

For the  \fomegaKs\ and \fetapKz\
analyses, we use large control samples to determine any needed adjustments to the
signal PDF shapes that have initially been determined from Monte Carlo.  For
\mes\ and \DE\ in \fomegaKs\ and \fetapKs, we use the decay $\Bm\ra\pim\Dz$ with $\Dz\ra K^{-}\pip\piz$,
which has similar topology to the modes under study here.  We select this
sample by making loose requirements on \mes and \DE, and requiring for
the $D^0$ 
candidate mass $1845<m_D<1885$ \mev.  We also 
place kinematic requirements on the $D$ and $B$ daughters to force the
charmed decay to look as much like that of a charmless decay as possible.
These selection criteria are applied both to the data and to MC.  For
\xf , we use a sample of \etaprgKp\ decays
selected with requirements very similar to those of our signal modes.  From
these control samples, we determine small adjustments (of the order of few \mev)
to the mean value of the signal \DE\ distribution.  The means and widths of the other
distributions do not need adjustment.   

For the $\omega$ mass line shape, we use $\omega$ production in the data
sidebands.  The means and resolutions of the invariant mass
distributions are compared between data and MC, and small adjustments
are made to the PDF parameterizations.  These studies also provide
uncertainties in the agreement between data and MC that are used for
evaluation of systematic errors.  For the \fpizKs\ analysis, we apply no
correction to the signal PDF shapes, but we evaluate the related systematic
error as described in Sec.~\ref{sec:syst}.

%%%%%%%%%%%%%%%%%%%%%%%%%%%%%%%%
% The Fits
%%%%%%%%%%%%%%%%%%%%%%%%%%%%%%%%%

\subsection{Fit variables}

For the \fomegaKs\ analysis we perform a fit with 25 free
parameters: $S$, $C$, signal yield, continuum background yield and
fractions (6), and the background PDF parameters for \dt, \mes, \DE, \xf,
$m_\omega$, and \hel\ (15).  For the five \fetapKs\ channels 
we perform a single fit with 98 free parameters:
$S$, $C$, signal yields (5), $\etapr_{\rho\gamma} \KS$ charm \BB\
background yields (2), continuum background yields (5) and fractions
(30), and the background PDF parameters for \dt, \mes, \DE, and \xf\ (54).  
Similarly the two \fetapKl\ channels are fit jointly, with $34$ parameters:
$S$, $C$, signal yields (2), background yields (2), fractions (12), and
PDF parameters (16).  For the
\fpizKs\ analysis we perform a fit with $36$ free parameters: $S$, $C$,
signal yield, the means of \mmiss\ and $m_{B}$ signal PDFs, background yield, background
PDF parameters for \dt, $m_B$, $m_{\rm miss}$, $L_2/L_0$, $\cos\theta_B$
($16$), background tagging efficiencies ($12$), and the fraction of good
events ($2$).  For the signal, charm \BB, and charmless \BB\ components the
parameters $\tau$ and $\deltamd$ are fixed to world-average values
\cite{PDG2006}; for the \BB\ components $S$ and $C$ are fixed to zero and
then varied to obtain the related systematic uncertainty as described below;
for the \qqbar\ model $\tau$ is fixed to zero.

\subsection{Fit validation}
\label{sec:Validation}

We test the fitting procedure by applying it to ensembles
of simulated experiments with \qqbar\ and \BB\ charmed events drawn from 
the PDF into which we have embedded the expected number of signal and 
\BB\ charmless background events (with the expected values of $S$ and
$C$) randomly extracted from the fully  
simulated MC samples.  We find biases (measured$-$expected) for
$\sksomega$, $\sksetap$, $\cksetap$, $\skletap$, and $\ckletap$ of 
$0.034\pm0.010$, $0.006\pm0.006$, $-0.008\pm0.005$, $-0.022\pm0.014$, 
and $-0.013\pm0.009$, respectively.  
These small biases are  
due to neglected correlations among the observables, 
contamination of the signal by self-crossfeed,  
and the small signal event yield in \fomegaKs.  We apply additive
corrections to the final results for these biases.  For $\cksomega$,
$\skspiz$, and $\ckspiz$ we make no correction but assign a systematic
uncertainty as described in Sec.~\ref{sec:syst}.

\section{Fit results}
\label{sec:results}

\begin{table}[!b]
\caption{Results of the fits.  Signal yields quoted here include events with
  no flavor tag information.  Subscripts for \etapr\ decay modes denote (1)
  \etapeppgg, (2) \etaprg, and (3) \etapeppppp.}  
\label{tab:Results}
\vspace*{-0.3cm}
\begin{center}
\begin{tabular}{lcccc}
\hline\hline
Mode                              &\# events &Signal yield  & $ -\eta_f S_f$       &         $C_f$      \\
\hline
$\fomegaKs$                   	&$17422$&$163\pm 18$  &$0.55^{+0.26}_{-0.29}$ &$-0.52^{+0.22}_{-0.20}$ \\
\hline
$\etapr_1 K^0_{\pi^+\pi^-}$	&$~1470$&$~472\pm 24$ &$0.70\pm0.17$ &$   -0.17\pm0.11$ \\
$\etapr_2 K^0_{\pi^+\pi^-}$	&$22775$&$1005\pm 40$ &$0.46\pm0.12$ &$   -0.13\pm0.09$ \\
$\etapr_3 K^0_{\pi^+\pi^-}$	&$~~513$&$~171\pm 14$ &$0.76\pm0.26$ &$\msp0.05\pm0.20$ \\
$\etapr_1 K^0_{\pi^0\pi^0}$	&$~1056$&$~105\pm 13$ &$0.51\pm0.34$ &$   -0.19\pm0.30$ \\
$\etapr_2 K^0_{\pi^0\pi^0}$	&$27057$&$~206\pm 28$ &$0.26\pm0.33$ &$\msp0.04\pm0.26$ \\
\hline
$\fetapKs$		        &$52871$&$1959\pm 58$ &$0.53\pm0.08$ &$   -0.11\pm0.06$ \\
\hline
$\etapr_1\KL$		        &$18036$&$386\pm 32$  &$0.75\pm0.22$ &$\msp0.02\pm0.16$ \\
$\etapr_3\KL$		        &$6213$&$169\pm 21$  &$0.87\pm0.30$ &$\msp0.19\pm0.25$ \\
\hline
$\fetapKl$		        &$24249$&$556\pm 38$ &$0.82^{+0.17}_{-0.19}$ &$\msp0.09^{+0.13}_{-0.14}$ \\ 
\hline
$\fpizKs$                       &$21412$&$556\pm 32$  &$0.55 \pm 0.20$   &$0.13 \pm 0.13$ \\
\hline\hline

\end{tabular}
\end{center}
\vspace*{-0.3cm}
\end{table}

\begin{figure}[!tb]
%\hspace*{-0.5cm}
\includegraphics[angle=0,width=0.48\textwidth]{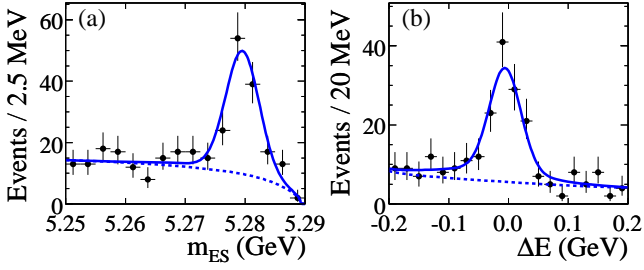}\\
 \caption{\label{fig:projMbDE_omks} Distributions for \fomegaKs\ projected (see text) onto
   (a) \mb\ and (b) \DE.  The solid
   lines show the fit result and the dashed lines show the
   background contributions.} 
\end{figure}

\begin{figure}[!b]
\includegraphics[angle=0,width=0.49\textwidth]{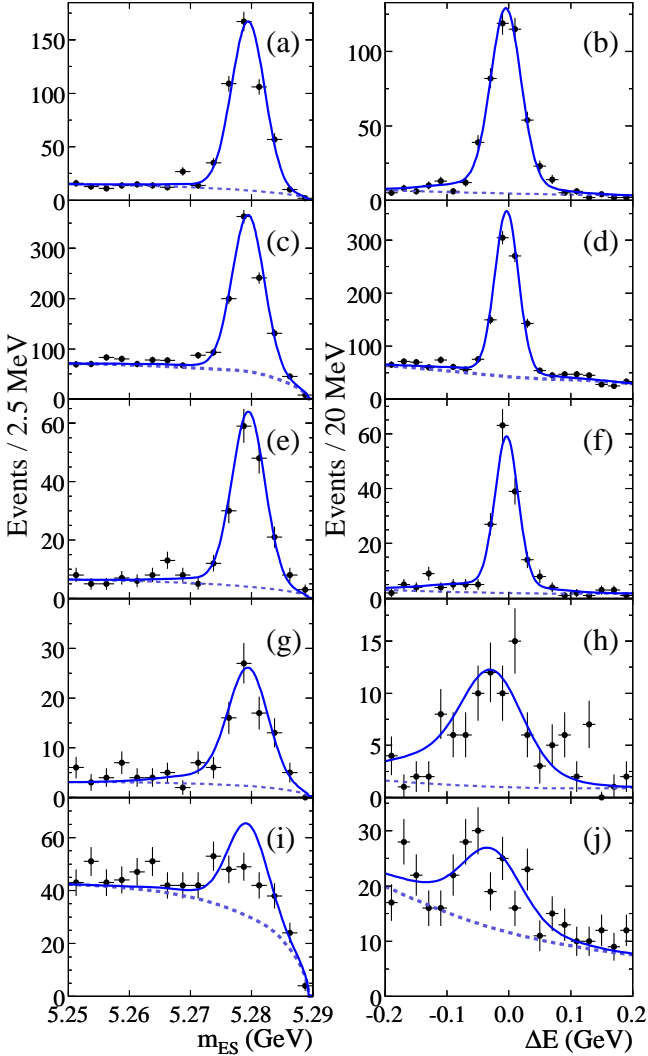}\\
 \caption{\label{fig:projMbDE} Distributions for 
(a,b) $\etapr_{\eta_{(\gamma\gamma)}\pi\pi} K^0_{\pi^+\pi^-}$, 
(c,d) $\etapr_{\rho\gamma} K^0_{\pi^+\pi^-}$, 
(e,f) $\etapr_{\eta_{(3\pi)}\pi\pi} K^0_{\pi^+\pi^-}$, 
(g,h) $\etapr_{\eta_{(\gamma\gamma)}\pi\pi} K^0_{\pi^0\pi^0}$, and
(i,j) $\etapr_{\rho\gamma} K^0_{\pi^0\pi^0}$ 
   projected (see text) onto (\mb,\,\DE).  The solid
   lines show the fit result and the dashed lines show the
   background contributions.} 
\end{figure}

\begin{figure}[!htb]
\includegraphics[angle=0,width=0.50\textwidth]{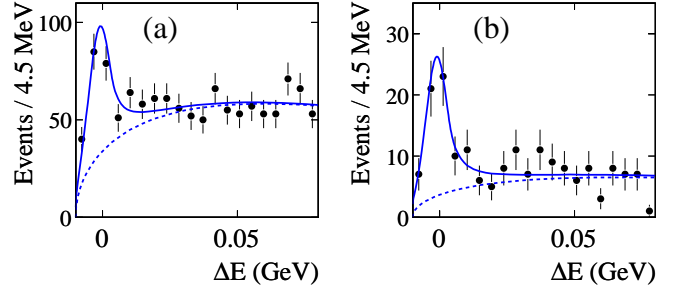}\\
 \caption{\label{fig:projKlDE} Distributions for (a) $\etapr_{\eta_{(\gamma\gamma)}\pi\pi} K^0_L$ and
   (b) $\etapr_{\eta_{(3\pi)}\pi\pi} K^0_L$ projected (see
   text) onto \DE.  The solid lines show the fit result and the dashed
   lines show the background contributions. }
\end{figure}

\begin{figure}[!htb]
  \begin{center}
    \includegraphics[width=0.4\textwidth]{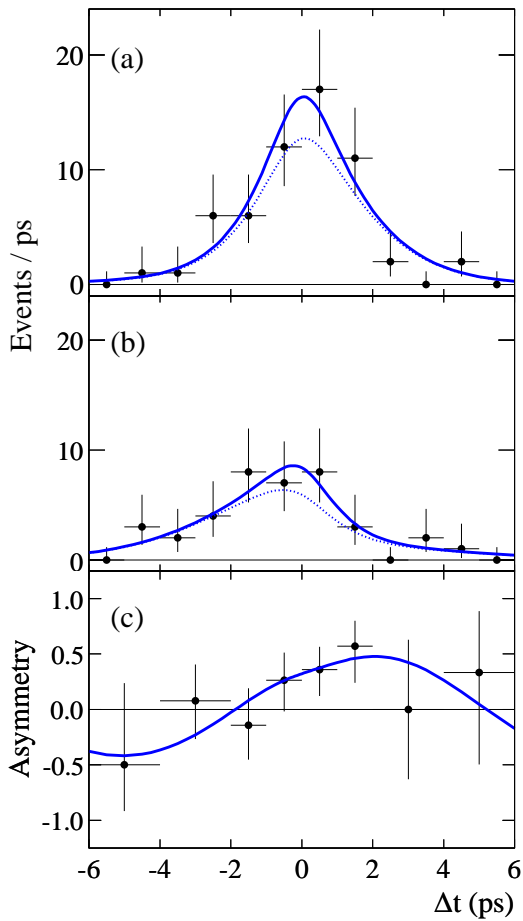}
\end{center}
  \vspace*{-0.5cm} 
  \caption{Data and model projections for \fomegaKs\ onto \dt\ for (a) \Bz\ 
    and (b) \Bzb\ tags.  We show data as points with error bars and the
    total fit function (total signal) with a the solid (dotted) line.
    In (c) we show the raw asymmetry,
    $(N_{\Bz}-N_{\Bzb})/(N_{\Bz}+N_{\Bzb})$ with a
    solid line representing
    the fit function.}
  \label{fig:DeltaTProj_omks}
\end{figure}

\begin{figure}[!htb]
  \begin{center}
    \includegraphics[width=0.4\textwidth]{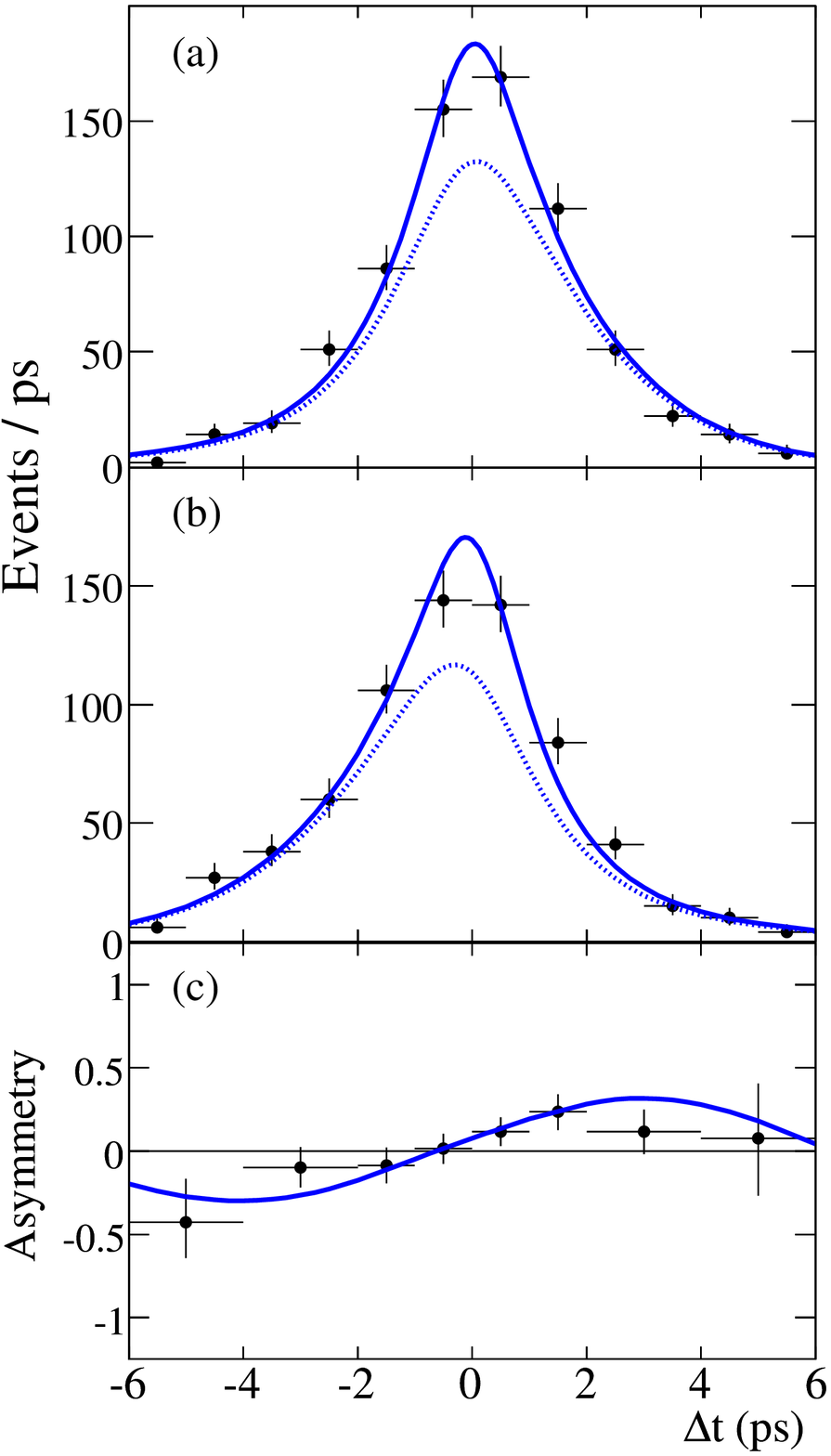}
\end{center}
  \vspace*{-0.5cm} 
  \caption{Data and model projections for \fetapKs\ onto \dt\ for (a) \Bz\ 
    and (b) \Bzb\ tags. Points with error bars represent the 
    data; the solid (dotted) line displays the total fit function 
    (total signal). In (c) we show the raw asymmetry,
    $(N_{\Bz}-N_{\Bzb})/(N_{\Bz}+N_{\Bzb})$; the solid line represents 
    the fit function.}
  \label{fig:DeltaTProj_etapks}
\end{figure}

\begin{figure}[!htb]
  \begin{center}
     \includegraphics[width=0.4\textwidth]{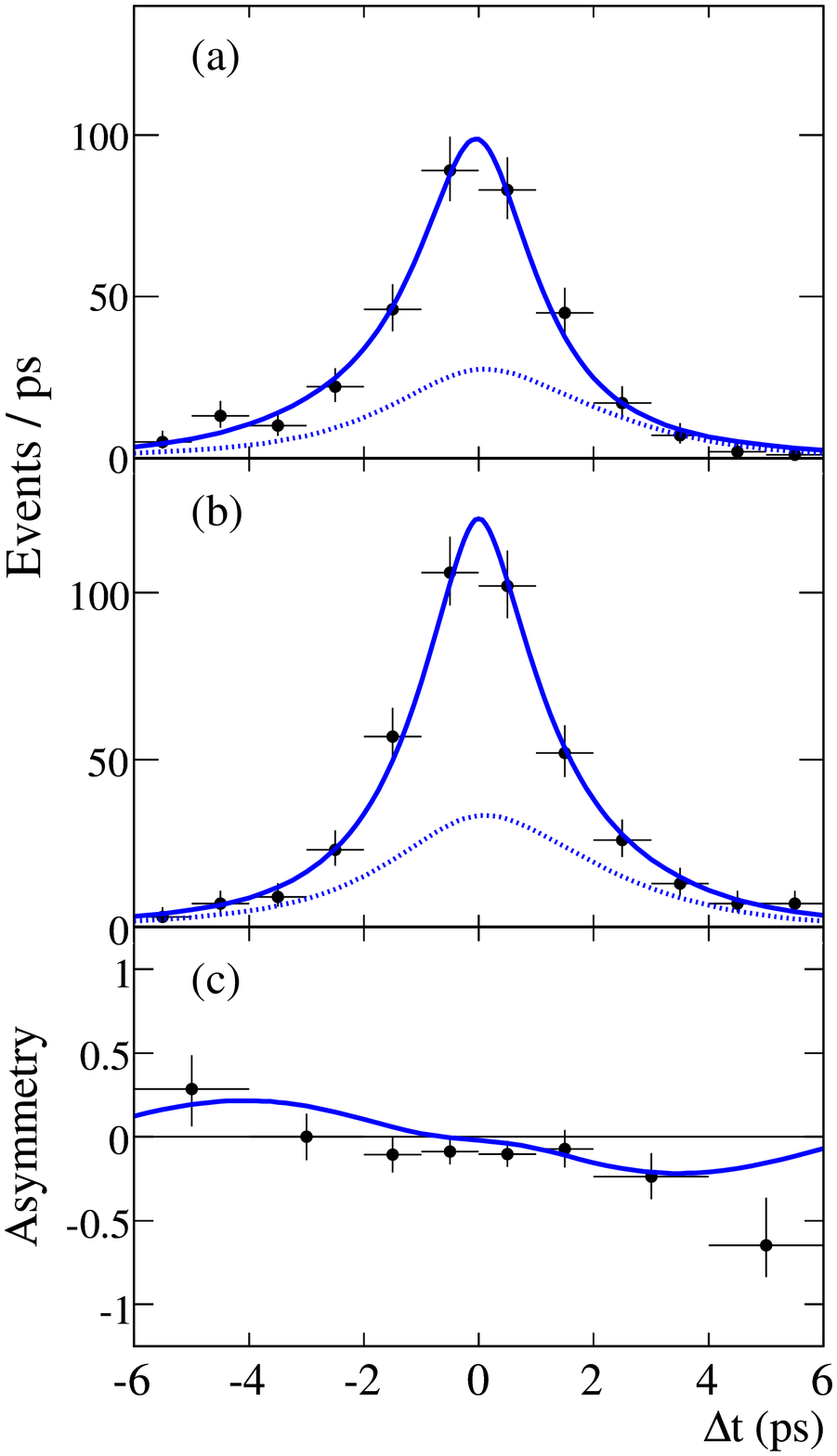}
\end{center}
  \vspace*{-0.5cm} 
  \caption{Data and model projections for \etapKl\ onto \dt\ for \Bz\ 
    (a) and \Bzb\ (b) tags. Points with error bars represent the 
    data; the solid (dotted) line displays the total fit function 
    (total signal). In (c) we show the raw asymmetry,
    $(N_{\Bz}-N_{\Bzb})/(N_{\Bz}+N_{\Bzb})$; the solid line represents 
    the fit function.}
  \label{fig:DeltaTProj_etapkl}
\end{figure}

\begin{figure}[!htb]
  \begin{center}
    \includegraphics[width=0.43\textwidth]{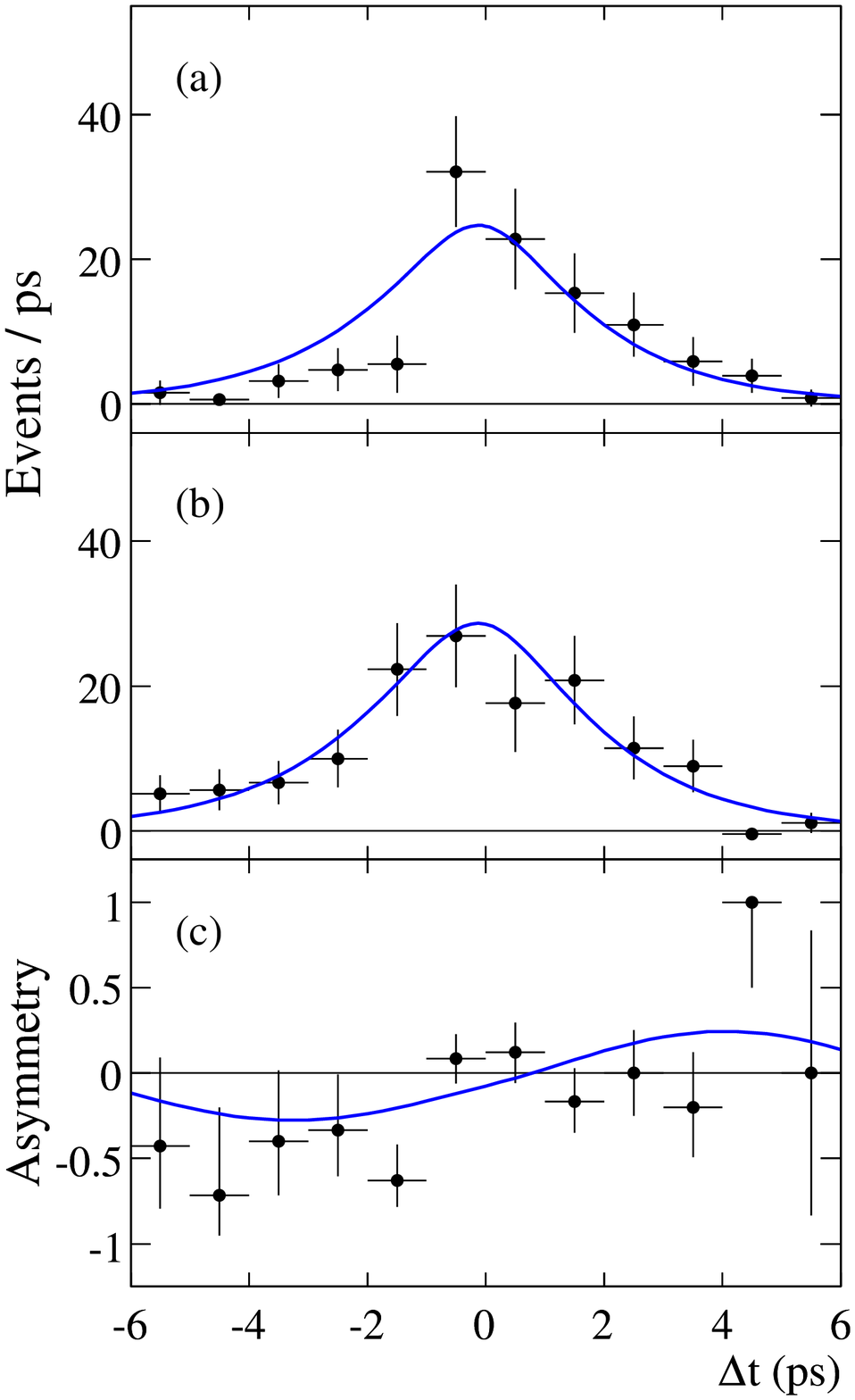}
\end{center}
  \vspace*{-0.5cm} 
  \caption{Data and model projections for \fpizKs\ onto \dt\ for (a) \Bz\
    and (b) \Bzb\ tags. Points with error bars represent the
    signal where backgroud is subtracted using an event weighting
    techinque~\cite{ref:splots}; the solid line displays the signal fit
    function. In (c) we show the raw asymmetry,
    $(N_{\Bz}-N_{\Bzb})/(N_{\Bz}+N_{\Bzb})$; the solid line represents
    the fit function.}
  \label{fig:DeltaTProj_pizks}
\end{figure}

Results from the fits for the signal yields and the \CP parameters $\Sf$
and $\Cf$ are presented in Table \ref{tab:Results}.  
In Figs.\
\ref{fig:projMbDE_omks}--\ref{fig:DeltaTProj_pizks},
we show projections onto the kinematic variables and \deltat\ for subsets of the data for which the
ratio of the likelihood to be signal and the sum of likelihoods to be 
signal and background (computed without the variable plotted) exceeds
a mode-dependent threshold that optimizes the statistical significance of
the plotted signal.  In \fomegaKs\ the fraction of signal events with
respect to the total after this requirement 
has been applied is $\sim 70\%$, while in \fetapKs\ and \fetapKl, the
fraction of signal events is in the  $(42-85)\%$ and $(22-55)\%$ range
respectively, depending on the decay mode. 
In Fig.~\ref{fig:projMbDE_omks} we show the projections onto \mb\ and \DE\ 
for the \fomegaKs\ analysis; in Fig.~\ref{fig:projMbDE}\ we show the
projections onto \mb\ and \DE\ for \fetapKs; in Fig.~\ref{fig:projKlDE}
we show the \DE\ projections for \fetapKl.  The corresponding
information for \fpizKs\ is conveyed by the background-subtracted
distributions for $m_B$ and $m_{\rm miss}$ in Fig.~\ref{fig:kspi0splots}.

In Figs.~\ref{fig:DeltaTProj_omks}--\ref{fig:DeltaTProj_pizks},
we show the $\Delta t$ projections and the asymmetry 
$(N_{\Bz}-N_{\Bzb})/(N_{\Bz}+N_{\Bzb})$ for each final state.
In the $\omega\KS$, \fetapKs, \fetapKl,
and $\piz\KS$ analyses, we measure the correlation between $\Sf$ and $\Cf$ in
the fit to be 2.9\%, 
3.0\%, 1.0\%
and $-6.2$\%, respectively. 

\subsection{Crosschecks}
\label{sec:Crosschecks}
We perform several additional crosschecks of our analysis technique
including time-dependent fits for $B^\pm$ decays to the final
states $\etapr_{\eta(\gamma\gamma)\pi\pi}\Kpm$, $\etapr_{\rho\gamma}\Kpm$, and
$\etapr_{\eta(3\pi)\pi\pi}\Kpm$ in which measurements of $S$ and $C$ are
consistent with zero.
There are only small changes in the results when we do any of the following: 
fix $C =0$ or $S=0$, allow $S$ and $C$ to be different for each tagging 
category, remove each of the discriminating variables one by one, and 
allow the signal resolution model parameters to vary in the fit.

To validate the IP-constrained vertexing technique in \fpizKs,
we examine $\Bz\to\jpsi\KS$ decays in
data where $\jpsi\to\mup\mun$ or $\jpsi\to \epem$.  In these events
we determine \deltat\
in two ways: by fully reconstructing the
$\Bz$ decay vertex using the trajectories of charged daughters of the
$\jpsi$ and the $\KS$ mesons (standard method), or by neglecting the
$\jpsi$ contribution to the decay vertex and using the IP constraint
and the \KS{} trajectory only.  This study shows that within
statistical uncertainties 
of order 2\% of the error on \deltat,
the IP-constrained 
\deltat\ measurement is unbiased with respect to the standard technique and
that the fit values of $S_{\jpsi\KS}$ and $C_{\jpsi\KS}$ are consistent
between the two methods.

\section{Systematic uncertainties}
\label{sec:syst}

A number of sources of systematic uncertainties affecting the fit 
values of $S_f$ and $C_f$ have been considered.

We vary the parameters of the signal PDFs that are kept fixed
in the fit within their uncertainty and take as systematic error the
resulting changes of $S_f$ and $C_f$. These parameters include
$\tau$ and $\Delta m_d$, the mistag parameters $w$ and $\Delta w$, the
efficiencies of each tagging category, the parameters of the
resolution model, and the shift and scale factors applied to the variables
related to the $B$ kinematics and event shape variables that serve to distinguish
signal from background.  The deviations of $S_f$ and $C_f$ for \fetapKs\ and
\fetapKl\ for variations of $\tau$ and \deltamd are less than $0.002$. 

For the \fpizKs\ channel as an additional systematic error
associated with the shape of the PDFs we also use the largest
deviation observed when the parameters of the individual PDFs are
free in the fit.

As a systematic uncertainty related to the fit bias on \Sf\ and \Cf\ we assign
the statistical uncertainty on the bias obtained from simulated experiments
during the fit validation.  As explained in Sec.~\ref{sec:Tagging}, we
obtain parameters of the signal resolution model from a fit to the
\bflav\ sample instead of from a fit to signal MC.  We evaluate the
systematic uncertainty of this approach with two sets of simulated experiments
that differ only in the values of resolution model parameters (one set with
parameters from the \bflav\ sample and one set with parameters from MC).  We
take the difference in the average $S_f$ and $C_f$ from these two sets of
experiments as the related systematic error.

We evaluate the impact of potential biases arising from
the interference of doubly Cabibbo-suppressed decays with
the Cabibbo-favored decays on the tag-side of the event \cite{dcsd} 
by taking into account realistic values of the ratio between the two 
amplitudes and the relative phases.  
For \fomegaKs\ and \fetapKz\ we
estimate using MC, published measurements, and theoretical predictions that
conservative ranges of the net values for \CP\ parameters in the \BB
background are $S=[0,0.2]$ and $C=\pm0.1$ for the charmless background and
$S=\pm0.1$ and $C=\pm0.1$ for the charm background.  
We perform a fit  in which we
fix the parameters to these values and take the difference in signal \CP
parameters between this fit and the nominal fit as the systematic error.

For the \fomegaKs\ and \fetapKz\ channels we also vary the amount of the charmless
\BB\ background by $\pm20\%$.  For \fpizKs\ we do not include a \BB\ background
component in the fit but we embed \BB\ background events in the data
sample and extract the peaking background from the observed change in
the yield.  We use this yield to estimate the change in $S$ and $C$ due to the
\CP\ asymmetry of the peaking background.   We also measure the systematic
error associated with the vertex reconstruction by varying within
uncertainties the parameters of the alignment of the SVT and the position
and size of the beam spot.

We quantify the effects of self-crossfeed events in the \fetapKz\ analysis.  
For \fetapKs\ we perform sets of simulated experiments in which we embed only correctly
reconstructed signal events and compare the results to the nominal simulated
experiments (Sec.~\ref{sec:Validation}) in which we embed both correctly and
incorrectly reconstructed signal events.  We take the difference as the
systematic uncertainty related to self-crossfeed.  For the \etapeppppp\KL
analysis, in which we include a self-crossfeed component in the fit, we
perform a fit in which we take parameter values for the self-crossfeed
resolution model from self-crossfeed MC events instead of the nominal
\bflav\ sample.  We take the difference of the results from this fit and the
nominal fit as the self-crossfeed systematic for \fetapKl.  The effects of
self-crossfeed are negligible for \fomegaKs\ and \fpizKs.

\begin{table*}[!ht]
\begin{center}
\caption{Summary of systematic uncertainties affecting $S_f$ and $C_f$.} 
\label{tab:syst_SC}
\begin{tabular}{lcc|cc|cc|cc}
\dbline
Source           & \sksomega\: & \cksomega\: & \sksetap\:  & \cksetap\:  & \skletap\:  & \ckletap\:  & \skspiz\:   & \ckspiz\:  \\
\sgline                                         
Variation of PDF parameters                     & 0.012 \: & 0.019 \: & \:0.006 \: & \:0.009\: &  \:0.009  & \:0.007  & \:0.010\: & \:0.012\: \\
Bias correction                                 & 0.010 \: & 0.007 \: & \:0.006 \: & \:0.005\: &  \:0.014  & \:0.009  & \:0.011\: & \:0.001\: \\
Interference from DCSD on tag side\hspace{3mm}  & 0.001 \: & 0.015 \: & \:0.001 \: & \:0.015\: &  \:0.001  & \:0.015  & \:0.001\: & \:0.015\: \\
\BB\ background                                 & 0.009 \: & 0.010 \: & \:0.009 \: & \:0.005\: &  -  & -  & \:0.005\: & \:0.001\: \\
Signal \dt\ parameters from \bflav\             & 0.002 \: & 0.001 \: & \:0.009 \: & \:0.015\: &  \:0.004  & \:0.008  & \:0.016\: & \:0.011\: \\
SVT alignment                                   & 0.011 \: & 0.003 \: & \:0.002 \: & \:0.003\: &  \:0.004  & \:0.004  & \:0.009\: & \:0.009\: \\
Beam-spot position and size                     & 0.000 \: & 0.000 \: & \:0.002 \: & \:0.001\: &  \:0.004  & \:0.003  & \:0.004\: & \:0.002\: \\
Vertexing method                                &  -       &   -      &    -       &  -        &  -  & -  & \:0.008\: & \:0.016\:\\
Self-crossfeed                                  &     -    &     -    & \:0.004 \: & \:0.001\: &  \:0.001  & \:0.004  &    -      &     -     \\ 
\sgline
Total           & 0.021 \: & 0.028 \: & \: 0.016 \: & 0.024 \: & \: 0.018 \: & 0.021 \: & \: 0.025 \: & 0.028 \: \\
\dbline
\end{tabular}
\vspace{-5mm}
\end{center}
\end{table*}

Finally, for the \fpizKs\ analysis we examine large samples of
simulated \Bztokspiz\ and \Bztojpsiks\ decays to quantify the
differences between resolution function parameter values obtained from
the \bflav\ sample and those of the signal channel; we use these
differences to evaluate uncertainties due to the use of the resolution
function extracted from the \bflav\ sample.  We also use the
differences between resolution function parameters extracted from data
and MC in the \Bztojpsiks\ decays to quantify possible problems in the
reconstruction of the $\KS$ vertex. We take the sum in quadrature of
these errors  
as the systematic error related to the vertexing method.

The contributions of the above sources of systematic uncertainties
to $S_f$ and $C_f$ are summarized in Table \ref{tab:syst_SC}.

\section{\boldmath $S$ and $C$ parameters for \etapKz}
\label{sec:combine}

As noted in Sec.\ \ref{sec:intro}, the final states \fetapKs\ and
\fetapKl\ have opposite \CP\ eigenvalues, and in the SM, if
$\Delta\Sf=0$, then $-\eta_f\Sf= \stwob$.  We therefore compute the
values of \skzetap\ and \ckzetap\ from our separate measurements with
\etapKs\ and \etapKl, taking $-\skletap$ in combination with \sksetap,
and \ckletap\ with \cksetap.

To represent the results of the individual fits, we project the
likelihood by maximizing ${\cal L}$ (Sec.\
\ref{sec:MLfit}) at a succession of fixed values of \Sf\ to obtain
${\cal L}(-\eta_f\Sf)$.  We then convolve this likelihood with a
Gaussian function representing the independent systematic errors for each mode.
The product of these convolved one-dimensional likelihood functions for the
two modes, shifted in $-\eta_f S_f$ by their respective corrections (Sec.\
\ref{sec:Validation}), gives the joint likelihood for \skzetap.  The
likelihood for \ckzetap\ is computed similarly.  Since the measured
correlation between \Sf\ and \Cf\ is small in our fits (Sec.\
\ref{sec:results}), we extract the central values and total
uncertainties of these quantities from these one-dimensional likelihood
functions.  Applying the same procedure without the convolution over
systematic errors yields the statistical component of the error.  The
systematic component is then extracted by subtraction in quadrature from
the total error.

\section{Summary and discussion}

In conclusion, we have used samples of 
$121\pm 13$ \omegaKs, 
$1457\pm 43$ \etapKs, 
$416\pm29 $ \etapKl, and
$411\pm 24$ \pizKs\ 
flavor-tagged events to measure the time-dependent \CP violation
parameters 
\begin{eqnarray}
\sksomega &=& \msp\SomegaKs\,\nonumber\\
\cksomega &=& \ComegaKs\,\nonumber\\
\skzetap &=&\msp\SetapKz\,\nonumber\\
\ckzetap &=& \CetapKz\,\nonumber\\
\skspiz &=& \msp\SpizKs\,\nonumber\\
\ckspiz &=& \msp\CpizKs\,\nonumber
\end{eqnarray}
where the first errors are statistical and the second systematic.
These results are consistent with and supersede our previous measurements 
\cite{PreviousOmK,PreviousEtapK,PreviousPizK}; 
they are also consistent with the world average of \stwob\ measured in $\Bz\ra
J/\psi\KS$ \cite{PDG2006}.

\section{Acknowledgements}
We are grateful for the 
extraordinary contributions of our \pep2\ colleagues in
achieving the excellent luminosity and machine conditions
that have made this work possible.
The success of this project also relies critically on the 
expertise and dedication of the computing organizations that 
support \babar.
The collaborating institutions wish to thank 
SLAC for its support and the kind hospitality extended to them. 
This work is supported by the
US Department of Energy
and National Science Foundation, the
Natural Sciences and Engineering Research Council (Canada),
the Commissariat \`a l'Energie Atomique and
Institut National de Physique Nucl\'eaire et de Physique des Particules
(France), the
Bundesministerium f\"ur Bildung und Forschung and
Deutsche Forschungsgemeinschaft
(Germany), the
Istituto Nazionale di Fisica Nucleare (Italy),
the Foundation for Fundamental Research on Matter (The Netherlands),
the Research Council of Norway, the
Ministry of Education and Science of the Russian Federation, 
Ministerio de Educaci\'on y Ciencia (Spain), and the
Science and Technology Facilities Council (United Kingdom).
Individuals have received support from 
the Marie-Curie IEF program (European Union) and
the A. P. Sloan Foundation.

\end{document}